\documentclass[prd,showpacs,twocolumn,
amsmath,amssymb,superscriptaddress,floatfix,nofootinbib]{revtex4-1}
\usepackage[utf8]{inputenc}
\usepackage[sort&compress]{natbib}
\usepackage[normalem]{ulem}
\usepackage{lipsum} 
\usepackage{bm}
\usepackage{times}
\usepackage{amssymb,amsbsy,amsmath,amsfonts}
\usepackage{graphicx}
\usepackage{float}
\usepackage{slashed}
\usepackage{color}
\usepackage{morefloats}
\usepackage{rotating}
\usepackage{srcltx}
\usepackage{slashed}
\usepackage{subfigure}
\usepackage{multirow}
\usepackage{verbatim}
\usepackage{hyperref}
\usepackage{tabularx}
\usepackage{adjustbox}
\usepackage[dvipsnames]{xcolor}
\usepackage{nicefrac}
\usepackage{tikz-feynman}
\allowdisplaybreaks  
\usetikzlibrary{decorations.pathmorphing} 

\usepackage{hyperref}
\hypersetup{
	colorlinks	=true,
	urlcolor	=blue,
	linkcolor	=blue,
	citecolor	=blue,
	pdftitle	={},
	pdfauthor	={Shi Rui-Xiang, Li-Sheng Geng},
	pdfsubject	={Two-pole structures demystified: chiral dynamics at work}
	}

\begin{document}

\title{Radiative decays of the $\Lambda(1520)$ as a dynamically generated resonance}

\author{Rui-Xiang Shi}
\affiliation{Department of Physics, Guangxi Normal University, Guilin 541004, China}
\affiliation{Guangxi Key Laboratory of Nuclear Physics and Technology, Guangxi Normal University, Guilin 541004, China}

\author{Yu-Bao Zhang}
\affiliation{Department of Physics, Guangxi Normal University, Guilin 541004, China}
\affiliation{Guangxi Key Laboratory of Nuclear Physics and Technology, Guangxi Normal University, Guilin 541004, China}

\author{Jun-Xu Lu}
\email[Corresponding author: ]{ljxwohool@buaa.edu.cn}
\affiliation{School of Physics, Beihang University, Beijing 102206, China}

\author{Chun-Yan Song}
\email[Corresponding author: ]{cysong@buaa.edu.cn}
\affiliation{School of Physics, Beihang University, Beijing 102206, China}

\author{Li-Sheng Geng}
\email[Corresponding author: ]{lisheng.geng@buaa.edu.cn}
\affiliation{School of Physics,  Beihang University, Beijing 102206, China}
\affiliation{Sino-French Carbon Neutrality Research Center, \'Ecole Centrale de P\'ekin/School of General Engineering, Beihang University, Beijing 100191, China}
\affiliation{Peng Huanwu Collaborative Center for Research and Education, Beihang University, Beijing 100191, China}
\affiliation{Southern Center for Nuclear-Science Theory (SCNT), Institute of Modern Physics, Chinese Academy of Sciences, Huizhou 516000, China}

\begin{abstract}

Inspired by the latest BESIII measurement of the $\Lambda(1520)\to\gamma\Sigma^0$ radiative decay, we systematically study the decays $\Lambda(1520)\to\gamma\Lambda(\Sigma^0)$ within the chiral unitary approach, where the $\Lambda(1520)$ is treated as a dynamically generated resonance from meson-baryon interactions. Compared with previous chiral unitary studies, we adopt dimensional regularization for $S$-wave loop integrals to preserve gauge invariance and, for the first time, include Feynman diagrams with photon coupling to intermediate baryons. Our calculated partial decay width $\Gamma(\Lambda(1520)\to\gamma\Sigma^0)$ agrees well with the new BESIII data, whereas the predicted $\Gamma(\Lambda(1520)\to\gamma\Lambda)$ is considerably smaller than the CLAS experimental result. By comparing our results with predictions from various quark models, together with an analysis of the $\Lambda(1520)$ compositeness, we discuss the internal nature of the $\Lambda(1520)$ resonance, highlight its complex component structure, and stress the need for more refined theoretical frameworks and further experimental measurements.

\end{abstract}

\maketitle

\section{Introduction}
Electromagnetic radiative decays of excited hyperons serve as a sensitive probe for investigating hadron structure, particularly in resolving the ambiguous nature of certain resonances that may either predominantly consist of conventional three-quark configurations or are more likely dynamically generated meson-baryon molecular states~\cite{Doring:2025wms,Doring:2025sgb}. For half a century, such decays have been investigated within various theoretical frameworks, such as the chiral unitary approach~\cite{Geng:2007hz,Doring:2007rz,Doring:2006ub,Doring:2010rd,Shen:2025xcq}, quark models~\cite{Yu:2006sc,VanCauteren:2005sm,Darewych:1983yw,Kaxiras:1985zv,Warns:1990xi,Koniuk:1979vy,Capstick:2000qj,An:2010wb,Umino:1991dk}, and other approaches~\cite{Kim:2021wov,Salone:2021bvx,Bijker:2000gq}. In contrast to their theoretical counterparts, experimental measurements of the radiative decays of excited hyperon states remain limited, primarily due to their small branching fractions~\cite{CLAS:2005bgo,Bertini:1987ye,Mast:1968ltv,SPHINX:2004vuw,l6g2-2wg6,ParticleDataGroup:2024cfk}. Notably, reconciling theoretical descriptions with experimental results for radiative decays of hyperon resonances has remained a long-standing and significant challenge. 

Very recently, the BESIII Collaboration~\cite{l6g2-2wg6} reported new measurements of  the $\Lambda(1520)\to\gamma\Sigma^0$ and the ratio of the partial decay widths of the $\Lambda(1520)$, which are 
\begin{eqnarray}
&&\Gamma(\Lambda(1520)\to\gamma\Sigma^0)=46.8(66)(93){\rm ~keV},\\
&&{\cal R}_{\Lambda(1520)}=\frac{\Gamma(\Lambda(1520)\to\gamma\Lambda)}{\Gamma(\Lambda(1520)\to\gamma\Sigma^0)}=3.19(34)(19),
\end{eqnarray}
where the uncertainties in parentheses correspond to statistical and systematic errors, respectively. Note that the $\Lambda(1520)$ radiative decay has been measured previously~\cite{CLAS:2005bgo,Bertini:1987ye,Mast:1968ltv,SPHINX:2004vuw}. Except for the predictions from a non-relativistic quark model~\cite{Kaxiras:1985zv} 40 years ago, non existing theoretical results can reproduce the BESIII data. As a result, we revisit the radiative decays of the $\Lambda(1520)$ resonance in this work. 

Although quark models~\cite{Yu:2006sc,VanCauteren:2005sm,Darewych:1983yw,Kaxiras:1985zv,Warns:1990xi,Koniuk:1979vy,Umino:1991dk} have been used to calculate the radiative decays $\Lambda(1520)\to\gamma\Lambda(\Sigma^0)$, we adopt the chiral unitary approach in this work because the state can contain substantial or even dominant meson-baryon configurations~\cite{Sarkar:2005ap,Sarkar:2004jh,Roca:2006sz,Geng:2008er,Aceti:2014wka}. The chiral unitary approach is the unitary
extension of chiral perturbation theory~(U$\chi$PT) and can deal with hadron interactions at energies higher than the domain of chiral perturbation theory. Within this method, some well-known
resonances are dynamically generated in the coupled-channel formalism.  In the literature, three
schemes were proposed to unitarize coupled-channel amplitudes, i.e., the inverse amplitude method~\cite{Dobado:1996ps,Oller:1998hw}, the
N/D method~\cite{Oller:1998zr}, and the Bethe-Salpeter method~\cite{Oller:1997ti}. At higher energies, the discrepancies among the three methods prove to be minute when resonances are
already dynamically generated using the lowest-order interaction kernels. A summary and comparison of these methods can be found in Ref.~\cite{Oset:2006wz}. Radiative decays of the $\Lambda(1520)$ resonance have been studied within the coupled-channel formalism built upon the Bethe-Salpeter method~\cite{Doring:2006ub}. In this picture, the $J^P=\frac{3}{2}^-$ $\Lambda(1520)$ resonance is dynamically generated from the rescattering of the $\frac{3}{2}^+$ decuplet baryons with the $0^-$ pseudoscalar mesons. In Ref.~\cite{Doring:2006ub}, the authors used a cutoff scheme for the regularization of the $\Lambda(1520)\to\gamma\Lambda(\Sigma^0)$ Feynman diagrams. It comes at the cost of gauge invariance and inevitably leads to cutoff-dependent results. Furthermore, the
diagrams with a photon coupling to an intermediate baryon are not considered. Whether these shortcomings can explain the discrepancy between their results~\cite{Doring:2006ub} and the latest BESIII data needs to be examined, which is the main purpose of the present work.

In this work, we revisit the radiative decays $\Lambda(1520)\to\gamma\Lambda(\Sigma^0)$ by extending the formalism developed in Ref.~\cite{Doring:2006ub}. 
The main novelties of the present work are threefold: (i) We employ dimensional regularization for $S$-wave loop amplitudes, maintaining exact gauge invariance and avoiding the gauge-breaking cutoff scheme adopted in earlier studies. (ii) We include additional Feynman diagrams with photon coupling to intermediate octet baryons, which were neglected in previous chiral unitary calculations. (iii) We perform a detailed channel-by-channel and diagram-by-diagram decomposition, revealing the destructive interference mechanism that strongly suppresses \(\Lambda(1520)\to\gamma\Lambda\), and demonstrate the non-negligible role of $D$-wave contributions.

This work is organized as follows. In Sec.~\ref{SecII}, we briefly outline the description of the chiral unitary coupled-channel approach, and calculate the relevant Feynman diagrams. We show the numerical results for the radiative decays $\Lambda(1520)\to\gamma\Lambda$ and $\Lambda(1520)\to\gamma\Sigma^0$ in Sec.~\ref {SecIII}. A brief conclusion is given in Sec.~\ref{SecIV}.

\section{Theoretical framework}\label{SecII}

In this section, we introduce the chiral unitary coupled-channel approach and derive
the amplitudes for the radiative decays $\Lambda(1520)\to\gamma\Lambda$ and $\Lambda(1520)\to\gamma\Sigma^0$.

\subsection{Description of the $\Lambda(1520)$ state
in the chiral unitary coupled-channel
approach}

We explain how the $\Lambda(1520)$ is dynamically generated in a coupled-channel formalism. The lowest order Lagrangian describing the $S$-wave
interaction of the $\frac{3}{2}^+$ baryon decuplet with the $0^-$ meson
octet reads~\cite{Sarkar:2005ap,Sarkar:2004jh}
\begin{eqnarray}
{\cal L}=-i\bar{B}^{*\mu}\slashed{D}B_\mu^*,\label{Lag:TTMM}
\end{eqnarray}
where $B_{abc}^{*\mu}$ denotes the $\frac{3}{2}^+$ baryon decuplet field and the covariant derivative $D^\nu$ is given by
\begin{eqnarray}
D^\nu B_{abc}^{*\mu}=\partial^\nu B_{abc}^{*\mu}+\left(\Gamma^\nu,B^{*\mu}\right)_{abc},
\end{eqnarray}
with the following definitions
\begin{eqnarray}
&&\left(\Gamma^\nu,B^{*\mu}\right)_{abc}=\left(\Gamma^\nu\right)_a^d B_{dbc}^{*\mu}+\left(\Gamma^\nu\right)_b^d B_{adc}^{*\mu}+\left(\Gamma^\nu\right)_c^d B_{abd}^{*\mu},\nonumber\\
&&\Gamma^\nu=\frac{1}{2}\left(u^\dag\partial^\nu u+u\partial^\nu u^\dag\right).
\end{eqnarray}
Here $a$, $b$, and $c$ are the $SU(3)$ indices. $u={\rm exp}\left[i\Phi/\sqrt{2}F_\phi\right]$ with the unimodular matrix containing the pseudoscalar nonet $\Phi$, and $F_\phi$ is the
pseudoscalar decay constant. 

From the Lagrangian in Eq.~(\ref{Lag:TTMM}), one can obtain the interaction kernel $V$ for the $S$-wave transition amplitudes. The $V$ matrix is explicitly given in Refs.~\cite{Doring:2006ub,Sarkar:2005ap}. For the $\frac{3}{2}^-$ states with strangeness $S=-1$ and isospin $I=0$, the $S$-wave coupled channels are $\pi\Sigma^*$ and $K\Xi^*$. In this work, appropriate isospin projections are done as follows, where we have adopted the same isospin phase convention as Refs.~\cite{Doring:2006ub,Nieves:2001wt}:
\begin{eqnarray}
&&\left|\pi\Sigma^*,I=0\rangle\right.=-\frac{1}{\sqrt{3}}\left|\pi^+\Sigma^{*-}\rangle\right.-\frac{1}{\sqrt{3}}\left|\pi^0\Sigma^{*0}\rangle\right.\nonumber\\
&&\qquad\qquad\qquad~~~+\frac{1}{\sqrt{3}}\left|\pi^-\Sigma^{*+}\rangle\right.,\nonumber\\
&&\left|K\Xi^*,I=0\rangle\right.=\frac{1}{\sqrt{2}}\left|K^+\Xi^{*-}\rangle\right.-\frac{1}{\sqrt{2}}\left|K^0\Xi^{*0}\rangle\right..\label{Eq:IsoCCMBs}
\end{eqnarray}

The matrix $V$ is then used as the kernel for the Bethe-Salpeter
equation to obtain the unitary transition matrix~\cite{Oset:1997it}. This results
in the matrix equation
\begin{eqnarray}
T=V+VGT\Longrightarrow T=(1-VG)^{-1}V,\label{Eq:BS}
\end{eqnarray}
where the meson-baryon
loop function $G$ is a diagonal matrix and is given by~\cite{Sarkar:2005ap,Roca:2006sz,Doring:2006ub}
\begin{eqnarray}
&&G_{ii}=i2M_i\int\frac{d^4q}{(2\pi)^4}\frac{1}{(P-q)^2-M_i^2+i\epsilon}\nonumber\\
&&\qquad\times\frac{1}{q^2-m_i^2+i\epsilon}\left(\frac{|\vec{q}|}{|\vec{q}_{\rm on}|}\right)^{2l_{ii}},\label{Eq:Gl}  
\end{eqnarray}
in which $q$ is the four-momentum of the pseudoscalar
meson, and $P$ denotes the total four-momentum of the meson-baryon system. $M_i$ and $m_i$ stand for the masses of the baryons and mesons, respectively. $l_{ii}$ in the $S$- and $D$-wave channel is taken as 0 and 2, respectively. Here, $|\vec{q}|=\sqrt{(s-(m_i-M_i)^2)(s-(m_i+M_i)^2)}/(2\sqrt{s})$ with $s=P^2$, and $\vec{q}_{\rm on}=|\vec{q}|_{s=M_{\Lambda(1520)}^2}$.

\begin{figure}[htpb]
    \centering    \includegraphics[width=0.7\linewidth]{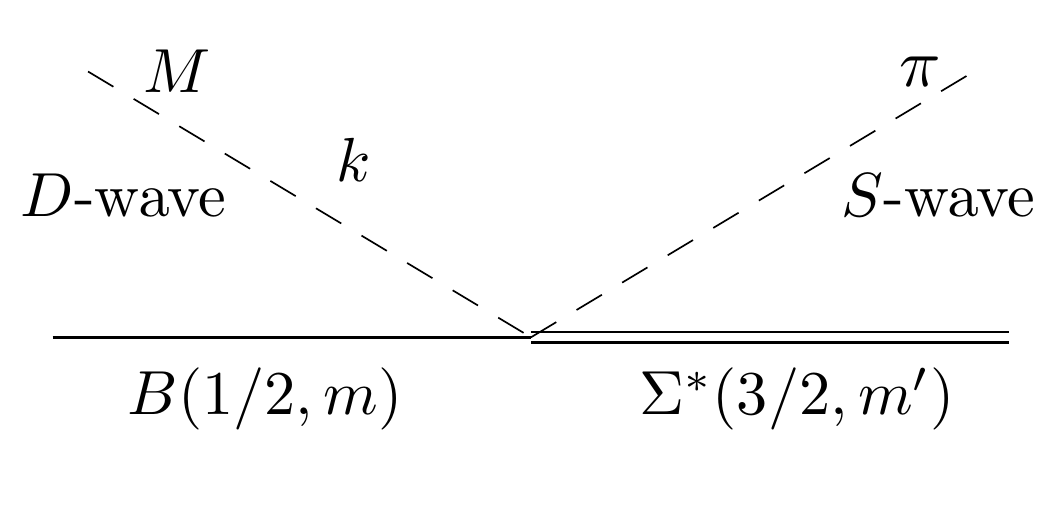}
    \caption{ The $MB\to\pi \Sigma^*$ vertex where $MB$ can be either $\bar{K}N$ or $\pi\Sigma$.}\label{Fig:dwaveAmp}
\end{figure}

Considering parity and angular momentum conservation, the $P$-wave case is forbidden. The $D$-wave channels $\bar{K}N$ and $\pi\Sigma$ couple to the $\Lambda(1520)$. These coupled channels can be represented as follows:
\begin{eqnarray}
&&\left|\bar{K}N,I=0\rangle\right.=\frac{1}{\sqrt{2}}\left|\bar{K}^0n\rangle\right.+\frac{1}{\sqrt{2}}\left|K^-p\rangle\right.,\nonumber\\
&&\left|\pi\Sigma,I=0\rangle\right.=-\frac{1}{\sqrt{3}}\left|\pi^+\Sigma^{-}\rangle\right.-\frac{1}{\sqrt{3}}\left|\pi^0\Sigma^{0}\rangle\right.\nonumber\\
&&\qquad\qquad\qquad~~~-\frac{1}{\sqrt{3}}\left|\pi^-\Sigma^{+}\rangle\right..\label{Eq:IsoCCMB}
\end{eqnarray}
For the purpose of calculating the loop integrations of the $\Lambda(1520)\to\gamma\Lambda(\Sigma^0)$, we have used the $\bar{K}N\to\pi\Sigma^*$ and $\pi\Sigma\to\pi\Sigma^*$ vertices as shown in Fig.~\ref{Fig:dwaveAmp}, which have the following form~\cite{Sarkar:2005ap,Doring:2006ub}
\begin{eqnarray}
&&-it_{\bar{K}N\to\pi\Sigma^*}=-i\beta_{\bar{K}N}\lvert\vec{q}\vert^2{\cal C}\left(\frac{1}{2}~2~\frac{3}{2};m,m^\prime-m\right)\nonumber\\
&&\qquad\qquad\qquad~\times Y_{2,m-m^\prime}(\hat{\vec{q}})(-1)^{m^\prime-m}\sqrt{4\pi},\label{Eq:dwaveVertex1}\\
&&-it_{\pi\Sigma\to\pi\Sigma^*}=-i\beta_{\pi\Sigma}\lvert\vec{q}\rvert^2{\cal C}\left(\frac{1}{2}~2~\frac{3}{2};m,m^\prime-m\right)\nonumber\\
&&\qquad\qquad\qquad~\times Y_{2,m-m^\prime}(\hat{\vec{q}})(-1)^{m^\prime-m}\sqrt{4\pi}.\label{Eq:dwaveVertex2}
\end{eqnarray}
Here $\beta_{\bar{K}N}$ and $\beta_{\pi\Sigma}$ represent the coupling strengths, and their values have to be determined via experiment~\cite{Roca:2006sz}. $Y_{2}(\hat{\bm{k}})$ is the spherical harmonic. $\vec{q}$ is the 3-momentum of the pseudoscalar meson. $m$ and $m^\prime$ denote the third components
of the spins of the initial octet baryon and the final $\Sigma^*$, respectively. ${\cal C}$ refers to the Clebsch-Gordan coefficient. In fact, the $K\Xi$ channel is also allowed in the $D$ wave. However, its influence should be small in the description of the $\Lambda(1520)$ resonance since the $K\Xi$ threshold is far from the region around 1520~{\rm MeV}. As a result, the $K\Xi$ channel has been safely ignored in our work. Considering the combination of $S$-wave and $D$-wave channels, the coupled channels in the kernel $V$ of the Bethe-Salpeter equation are in sequence $\pi\Sigma^*$, $K\Xi^*$, $\bar{K}N$, and $\pi\Sigma$ in the present work. The matrix $V$ containing 
the tree-level amplitudes is written as:
\begin{eqnarray}
 V = \begin{pmatrix}
C_{11}(k_1^0 + k_1^0) & C_{12}(k_1^0 + k_2^0) & \gamma_{13} q_3^2 & \gamma_{14} q_4^2 \\
C_{21}(k_2^0 + k_1^0) & C_{22}(k_2^0 + k_2^0) & 0 & 0 \\
\gamma_{13} q_3^2 & 0 & \gamma_{33} q_3^4 & \gamma_{34} q_3^2 q_4^2 \\
\gamma_{14} q_4^2 & 0 & \gamma_{34} q_3^2 q_4^2 & \gamma_{44} q_4^4
\end{pmatrix},~~~  
\end{eqnarray}
where $q_i$ is the on-shell three-momentum of the meson or baryon in the center-of-mass frame. The matrix elements $V_{11}$, $V_{12}$, $V_{21}$ and $V_{22}$ are derived from  the lowest 
order chiral Lagrangian in Eq.~(\ref{Lag:TTMM}). $V_{23}$ and $V_{24}$ involving the tree-level interaction of the $K\Xi^*$ channel to the $D$-wave channels are neglected in our work. This is because the $K\Xi^*$ threshold lies far above the $\Lambda(1520)$ mass, making its contribution negligible compared to that of the $\pi\Sigma^*$. For the same reason, the $K\Xi$ channel in the $D$ wave is also completely ignored. Five $D$-wave coupling strengths $\gamma_{ij}$ can be determined by fitting to $\bar{K}N\to\bar{K}N$ and $\bar{K}N\to\pi\Sigma$ data, as done in Ref.~\cite{Roca:2006sz}.

\subsection{Decay amplitudes of the $\Lambda(1520)\to\gamma\Lambda(\Sigma^0)$}

\begin{figure*}
    \centering
    \includegraphics[width=0.25\linewidth]{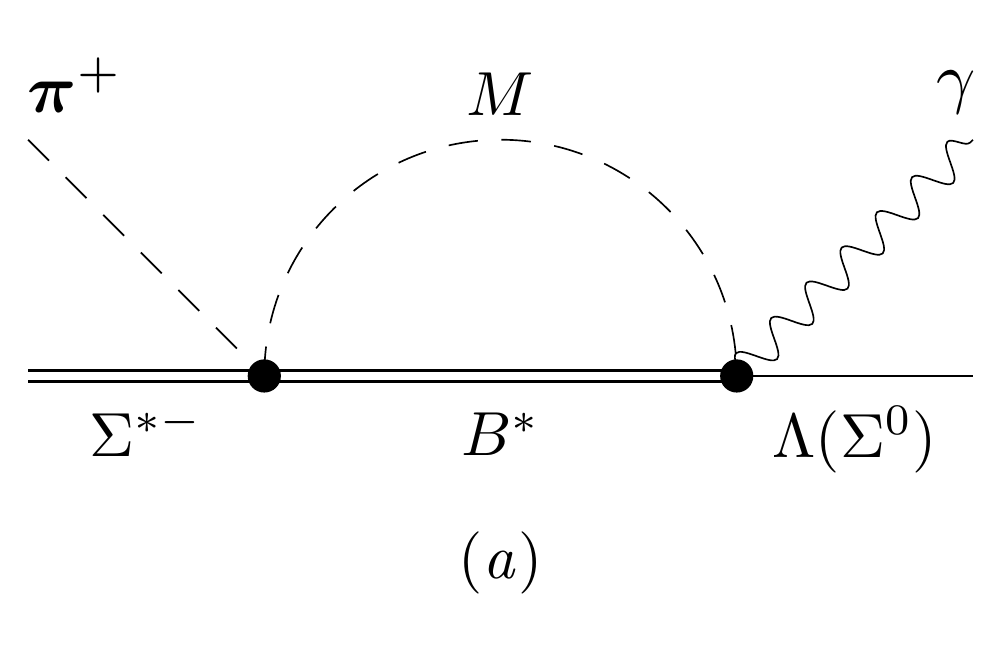}~\includegraphics[width=0.25\linewidth]{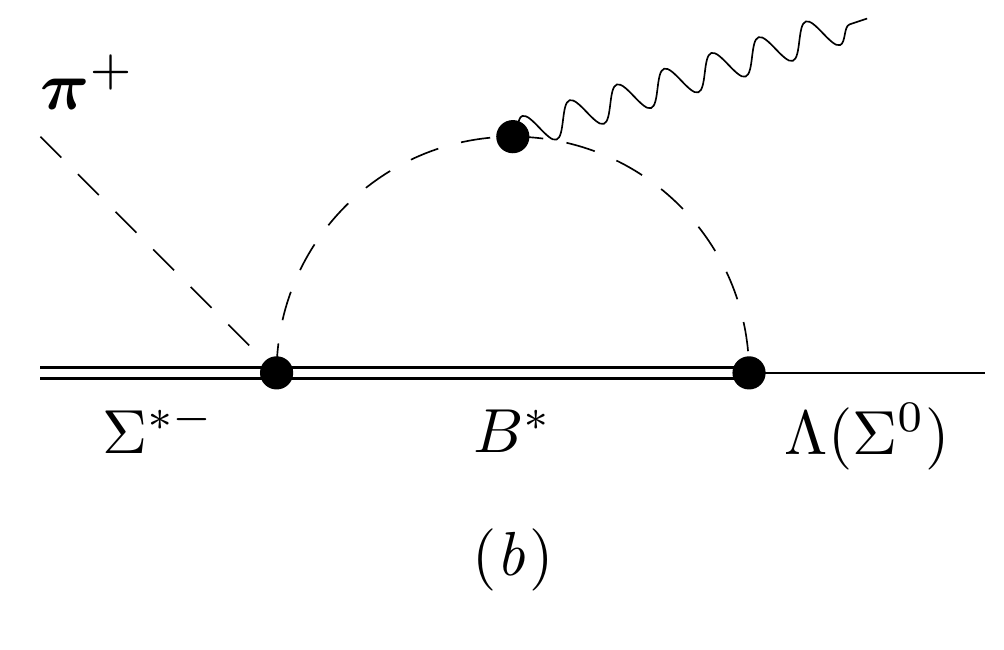}~\includegraphics[width=0.25\linewidth]{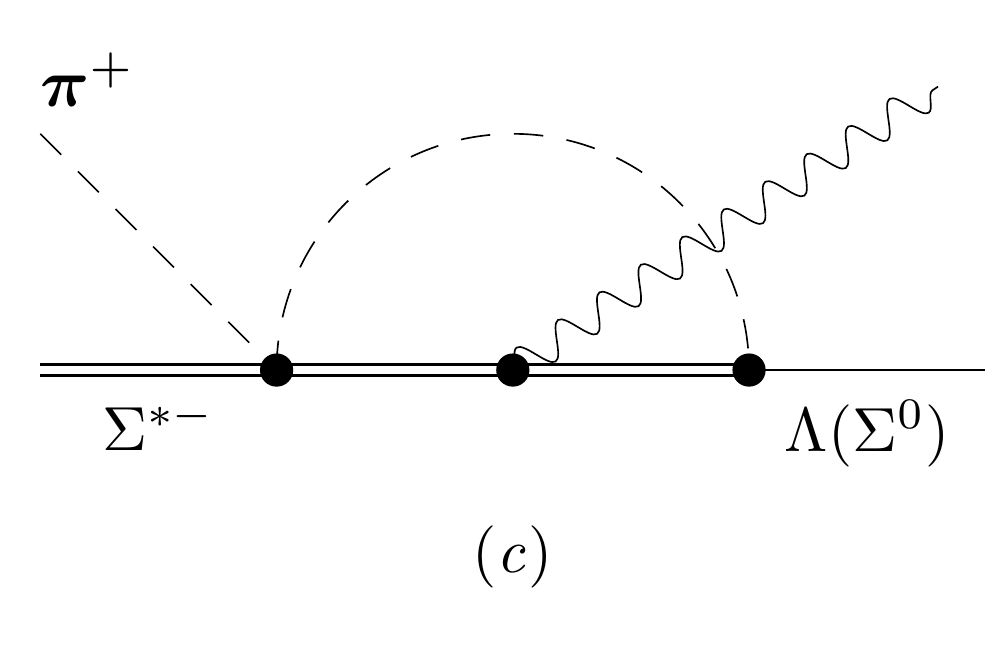}\\
    \includegraphics[width=0.25\linewidth]{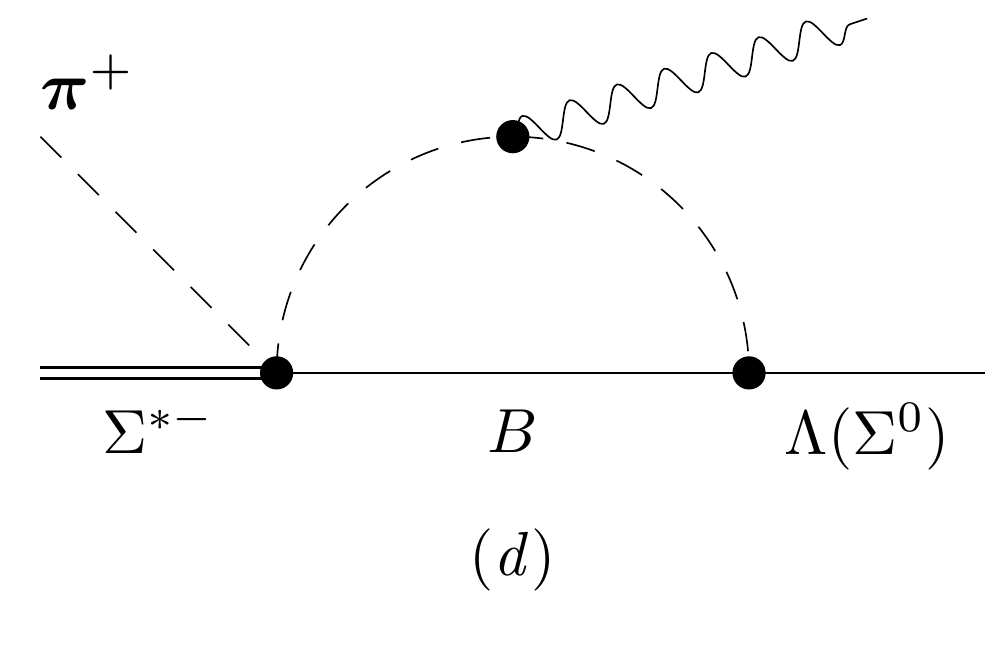}~\includegraphics[width=0.25\linewidth]{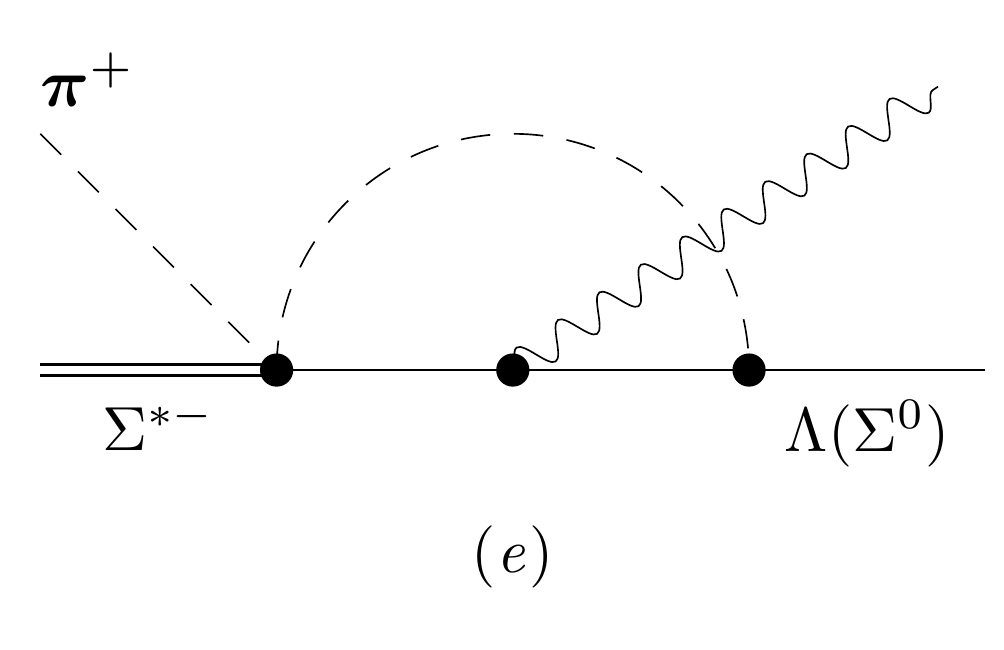}\\
    \caption{Radiative decay mechanism of the $\Lambda(1520)$ in $S$- and $D$-wave loops, where double-solid, solid, dashed, and wiggly lines represent decuplet baryons $(B^*)$, octet baryons $(B)$, Goldstone bosons $(M)$, and photons $(\gamma)$, respectively. Diagrams $(a)$, $(b)$, and $(c)$ show the couplings to the $MB^*$ loops in $S$ wave, which $MB^*$ can be the three coupled channels: $\pi^+\rm\Sigma^{*-}$, $\pi^-\rm\Sigma^{*+}$, and $K^+\rm\Xi^{*-}$. Diagrams $(d)$ and $(e)$ show these processes with $D$-wave couplings of $MB$~($K^-p$, $\pi^+\rm\Sigma^-$, and $\pi^-\rm\Sigma^+$) to the $\Lambda(1520)$.}\label{Fig:Lab1520}
\end{figure*}

In this subsection, we present how to compute the radiative decays $\Lambda(1520)\to\gamma\Lambda(\Sigma^0)$ as shown in Fig.~\ref{Fig:Lab1520}. The relevant loop diagrams are determined by
the lowest order Lagrangians ${\cal L}_B^{(1)}+{\cal L}_M^{(2)}+{\cal L}_{MB}^{(1)}+{\cal L}_{B^*}^{(1)}+{\cal L}_{B^*B}^{(1)}$~\cite{Geng:2007hz,Doring:2010rd,Xiao:2018rvd,Shi:2022dhw,Geng:2009ys,Geng:2009hh,Ren:2013dzt,Jenkins:1991es}, which are
\begin{eqnarray}
&&{\cal L}_B^{(1)}=\langle\bar{B}i\gamma^\mu D_\mu B-m_0\bar{B}B\rangle,\label{Eq:Lag16701}\\
&&{\cal L}_M^{(2)}=\frac{F_\phi^2}{4}\langle u_\mu u^\mu+\chi^+\rangle,\label{Eq:Lag16702}\\
&&{\cal L}_{MB}^{(1)}=\frac{D}{2}\langle\bar{B}\gamma^\mu\gamma^5\{u_\mu,B\}\rangle+\frac{F}{2}\langle\bar{B}\gamma^\mu\gamma^5\left[u_\mu,B\right]\rangle,\qquad\label{Eq:Lag16703}\\
&&{\cal L}_{B^*}^{(1)}=\bar{B}_{abc}^{*\mu}\left(i\gamma_{\mu\nu\alpha}D^\alpha-m_{B^*}\gamma_{\mu\nu}\right)B_{abc}^{*\nu},\label{Lag:15201}\\
&&{\cal L}_{B^*B}^{(1)}={\cal C}\left(\bar{B}^{*\mu}A_\mu B+\bar{B}A_\mu B^{*\mu}\right),\label{Lag:15202}
\end{eqnarray}
with
\begin{eqnarray}
&&D_\mu=\partial_\mu B+\left[\Gamma_\mu,B\right],~\chi^\pm=u^\dag\chi u^\dag\pm u\chi u,\nonumber\\
&&\Gamma_\mu=\frac{1}{2}\left(u^\dag\partial_\mu u+u\partial_\mu u^\dag\right)-\frac{i}{2}\left(u^\dag v_\mu u+uv_\mu u^\dag\right),\nonumber\\
&&u_\mu=i\left(u^\dag\partial_\mu u-u\partial_\mu u^\dag\right)+\left(u^\dag v_\mu u-uv_\mu u^\dag\right),\nonumber\\
&&A_\mu=\frac{i}{2}\left(u\partial_\mu u^\dag-u^\dag\partial_\mu u\right),
\end{eqnarray}
where $m_0$ denotes the baryon mass in the chiral limit~\cite{Ren:2012aj}, and $v_\mu$ is the vector source. $\chi=2B_0{\cal M}$ with the $B_0=|\langle0|\bar{q}q|0\rangle|$ and the quark mass matrix ${\cal M}={\rm diag}(m_q,m_q,m_s)$. $m_{B^*}$ is the decuplet baryon mass. The totally antisymmetric
gamma matrix products are defined as: $\gamma_{\mu\nu\alpha}=\frac{1}{2}\left\{\gamma_{\mu\nu},\gamma_\alpha\right\}$, and $\gamma_{\mu\nu}=\frac{1}{2}\left[\gamma_\mu,\gamma_\nu\right]$. The low energy constant ${\cal C}$ is determined from the strong decay of the $\Delta(1232)$, i.e., ${\cal C}=\frac{\sqrt{2}f_{\Delta\pi N}^*}{m_\pi}f_\pi$ with $f_{\Delta\pi N}^*=2.13$~\cite{Doring:2007rz,Doring:2006ub}, which characterizes the $\Delta\pi N$ coupling strength. In this work, as in Ref.~\cite{Geng:2008mf}, we take $F_\phi=1.17F_\pi$ with $F_\pi=92.4~{\rm MeV}$,  $D=0.80$, and $F=0.46$ for the axial and vector meson-baryon couplings. 

\begin{table}[h!]
 \caption{Coefficients of various $S$- and $D$-wave coupled channels in Fig.~\ref{Fig:Lab1520}.\label{Tab:Coeffi1520}}
\begin{center}
    \begin{tabular}{ccc}
      \hline
      \hline
      Coefficients  & $\Lambda(1520)\to\gamma\Lambda$ &  $\Lambda(1520)\to\gamma\Sigma^0$ \\
      \hline
      $\xi_{\pi^+\Sigma^{*-}}^{(s)}$ & $-\frac{\sqrt{6}}{2}$ & $-\frac{\sqrt{6}}{6}$\\
      
     $\xi_{\pi^-\Sigma^{*+}}^{(s)}$ & $-\frac{\sqrt{6}}{2}$ & $\frac{\sqrt{6}}{6}$\\
     
     $\xi_{K^+\Xi^{*-}}^{(s)}$ & $\frac{\sqrt{2}}{2}$ & $-\frac{\sqrt{6}}{6}$\\
     
    $\xi_{K^-p}^{(d)}$ & $\frac{1}{3}\left(D+3F\right)$ & $\frac{1}{\sqrt{3}}(F-D)$\\
    
    $\xi_{\pi^+\Sigma^-}^{(d)}$ & $\frac{2}{3}D$ & $\frac{2}{\sqrt{3}}F$\\
    
    $\xi_{\pi^-\Sigma^+}^{(d)}$ & $-\frac{2}{3}D$ & $\frac{2}{\sqrt{3}}F$\\
      \hline
      \hline
    \end{tabular}
  \end{center}
\end{table}

The amplitudes for the $\Lambda(1520)\to\gamma\Lambda(\Sigma^0)$ contain the contributions of the $S$-wave and $D$-wave channels which are represented by the diagrams $(a)$-$(c)$ and $(d)$-$(e)$ in Fig.~\ref{Fig:Lab1520}, respectively. Using the Lagrangians~(\ref{Eq:Lag16701})-(\ref{Lag:15202}) and summing over all diagrams, one obtains the corresponding amplitudes as follows:
\begin{eqnarray}
&& -it_{\gamma\Lambda(\Sigma^0)\to MB^{(*)}\to\pi^+\Sigma^{*-}}^{(I=0)}=-e\frac{f_{\Delta\pi N}^*}{m_\pi}S^\dag\cdot\epsilon\sum_{MB^*}\xi_{MB^*}^{(s)}\nonumber\\
&&\qquad\qquad\qquad\qquad\times\left(G_{MB^*}^{(2b)}+G_{MB^*}^{(2c)}\right)T_{MB^*\to\pi^+\Sigma^{*-}}\nonumber\\
&&\qquad\qquad\qquad\qquad+\frac{e}{F_\phi}S^\dag\cdot\epsilon\sum_{MB}\xi_{MB}^{(d)}\nonumber\\
&&\qquad\qquad\qquad\qquad\times\left(G_{MB}^{(2d)}+G_{MB}^{(2e)}\right)T_{MB\to\pi^+\Sigma^{*-}}.\qquad\label{Eq:Amp1520}
\end{eqnarray}
In Eq.~(\ref{Eq:Amp1520}), $MB^*$ and $MB$ are $\pi^+{\rm\Sigma}^{*-}$, $\pi^-{\rm\Sigma}^{*+}$, $K^+{\rm\Xi}^{*-}$, $K^-p$, $\pi^+{\rm\Sigma}^-$, and $\pi^-{\rm\Sigma}^+$ in sequence. $\xi_{MB^*}^{(s)}$ and $\xi_{MB}^{(d)}$ represent the flavor coefficients in the $S$ and $D$ waves, respectively, which are listed in Table~\ref{Tab:Coeffi1520}. It is to be noted that the Kroll-Ruderman structure analogous to diagram $(a)$ in Fig.~\ref{Fig:Lab1520} is not considered for the $D$-wave situation. This is because the combination of $S$-wave and $D$-wave vertices vanishes upon integrating over the loop momentum.

Next, we compute the loop functions. For all the diagrams in Fig.~\ref{Fig:Lab1520}, we use $P$, $k$, and $q$ to label the 4-momentum of the $\Lambda(1520)$, photon, and meson, respectively. Following Refs.~\cite{Roca:2006am,Doring:2007rz,Marco:1999df,Geng:2007hz}, the general expression of the $S$-wave loop functions can be written as
\begin{eqnarray}   G_{MB^*}S^\dag\cdot\epsilon=G_{MB^*}^{\mu\nu}S_\mu^\dag\epsilon_\nu,
\end{eqnarray}
with 
\begin{eqnarray}
G_{MB^*}^{\mu\nu}=ag^{\mu\nu}+bP^\mu P^\nu+cP^\mu k^\nu+dk^\mu P^\nu+ek^\mu k^\nu,~~\label{Eq:Gauge}
\end{eqnarray}
where $S_\mu^\dag$ is the spin $\frac{1}{2}\to\frac{3}{2}$ transition operator and $S_\mu^\dag=(0,\vec{S}^\dag)$. $\epsilon_\nu$ is the polarization vector of the photon. By employing the Lorentz condition $\epsilon_\nu k^\nu=0$ and the Ward identity $G_{MB^*}^{\mu\nu}k_\nu=0$ in Eq.~(\ref{Eq:Gauge}), one obtains the relation $a+d(P\cdot k)=0$. Furthermore, due to the fact $|\vec{P}|=0$ in the rest frame of the $\Lambda(1520)$ and the Coulomb gauge for the photon, i.e., $\epsilon^0=0$, only the term $ag^{\mu\nu}$ survives in Eq.~(\ref{Eq:Gauge}). In other words, the loop function $G_{MB^*}$ is equal to the coefficient $a$. However, we use the relation $a=-d(P\cdot k)$ to compute the coefficient $a$. The advantage of evaluating the coefficient $d$ is that the loop integral is finite and fewer terms contribute to it. One notes that diagram $(a)$ of Fig.~\ref{Fig:Lab1520} does not contribute to the coefficient $d$. As a result, the loop function $G_{MB^*}$ is only evaluated from the diagrams $(b)$ and $(c)$. To keep gauge invariance, the loop functions $G_{MB^*}^{(2b)}$ and $G_{MB^*}^{(2c)}$ are calculated using the dimensional regularization scheme, and their expressions correspond to the $G_{\rm g.i.}^{\rm I}$ and $G_{\rm g.i.}^{\rm II}$ in Ref.~\cite{Doring:2007rz} about the radiative decay of the $\Delta^*(1700)$. Notably, this way of dealing with the loop functions makes $G_{MB^*}^{(2b)}+G_{MB^*}^{(2c)}$ independent of the regularization scale $\mu$.

For the $D$-wave loop functions $G_{MB}^{(2d)}$ and $G_{MB}^{(2e)}$, their expressions are accompanied by the $|\vec{q}^2|$ factor from the $D$-wave vertices in Eqs.~(\ref{Eq:dwaveVertex1}) and (\ref{Eq:dwaveVertex2}). Due to the existence of the 3-momentum $\vec{q}$, it is not easy to treat the $D$-wave loop functions in the dimensional regularization scheme~\cite{Pavao:2018xub,Ikeno:2020vqv}. Instead, we use the cutoff scheme to evaluate $G_{MB}^{(2d)}$ and $G_{MB}^{(2e)}$. From Eq.~$(23)$ in Ref.~\cite{Doring:2006ub}, the loop function $G_{MB}^{(2d)}$ has been calculated in the cutoff regularization. Thus, this result is directly applicable. However, the loop function $G_{MB}^{(2e)}$ of a photon coupling to an intermediate baryon is evaluated for the first time. In the cutoff scheme, the $D$-wave loop function $G_{MB}^{(2e)}$ is given by 
\begin{eqnarray}
&&G_{MB}^{(2e)}=-i\int\frac{d^4q}{(2\pi)^4}\frac{|\vec{q}|^2}{q^2-m_M^2+i\epsilon}\frac{1}{P^0-q^0-E_B(\vec{q})+i\epsilon}\nonumber\\
&&\qquad\qquad\times\frac{1}{P^0-q^0-k^0-E_B^\prime+i\epsilon}\nonumber\\
&&\qquad\qquad\times\frac{m_B}{E_B(\vec{q})}\frac{m_B}{E_B^\prime(\vec{q}+\vec{k})}\frac{1}{2m_B}\frac{|\vec{q}|^2}{|\vec{q}_{MB}|^2}\nonumber\\
&&\qquad~~~=-\int \frac{\mathrm{d} |\vec{q}|}{(2\pi)^2} \frac{1}{P^0 - \omega_M - E_B(\vec{q}) + i\varepsilon} \nonumber\\
&&\qquad\qquad\times\frac{1}{P^0-\omega_M-k^0-E_B^\prime(\vec{q}+\vec{k})+i\varepsilon}\nonumber\\
&&\qquad\qquad\times \frac{m_B|\vec{q}|^4}{2\omega_ME_B(\vec{q})E_B^\prime(\vec{q}+\vec{k})}\frac{|\vec{q}|^2}{\left|\vec{q}_{MB}\right|^2},
\end{eqnarray}
where $E_B$ and $E_B^\prime$ are the energies of the octet baryon of mass $m_B$ at 
3-momentum $\vec{P}-\vec{q}$ and $\vec{P}-\vec{q}-\vec{k}$, respectively. $\omega_M$ denotes the energy of the pseudoscalar meson of mass $m_M$. $P^0$ and $k^0$ are the energies of the
on-shell $\Lambda(1520)$ and photon. $\vec{q}_{MB}$ is the on-shell 3-momentum of the pseudoscalar meson $(M)$ and octet baryon $(B)$ for the $\Lambda(1520)$ decay at rest. As done in Ref.~\cite{Doring:2006ub}, the cutoff values are obtained by matching the dimensional regularization and cutoff scheme of the meson-baryon $D$-wave loop at $\sqrt{s}=1520~{\rm MeV}$. This results in $\Lambda_{\bar{K}N}=507~{\rm MeV}$ and $\Lambda_{\pi\Sigma}=558~{\rm MeV}$, which are adopted in the present work.

\section{Results and discussions}\label{SecIII}

In this section, we first study the  $\Lambda(1520)\to\gamma\Lambda(\Sigma^0)$ decays within the chiral unitary approach.  Next, we compare the results for the $\Lambda(1520)\to\gamma\Lambda(\Sigma^0)$ decays obtained from the chiral unitary approach with those from various quark models, and analyze the compositeness of the $\Lambda(1520)$.

\subsection{Results for the $\Lambda(1520)\to\gamma\Lambda(\Sigma^0)$}

\begin{figure}[htpb]
    \centering
    \includegraphics[width=0.7\linewidth]{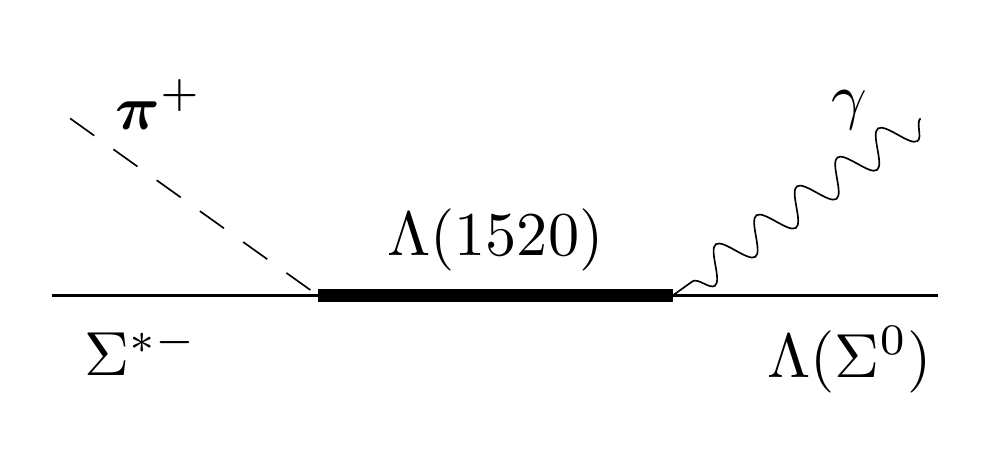}
    \caption{ Effective resonance representation of the $\Lambda(1520)\to\gamma\Lambda(\Sigma^0)$.}\label{Fig:EffLab1520}
\end{figure}

As can be seen in Eq.~(\ref{Eq:Amp1520}), the amplitudes for the radiative decays $\Lambda(1520)\to\gamma\Lambda(\Sigma^0)$ are written in terms of $T_{MB^*\to\pi^+\Sigma^{*-}}$ and $T_{MB\to\pi^+\Sigma^{*-}}$, the unitary solution of the Bethe-Salpeter
equation~(\ref{Eq:BS}) for meson-baryon scattering with the transitions from
channels $MB^*$ and $MB$ to the $\pi^+\Sigma^{*-}$ final state. $T_{MB^*\to\pi^+\Sigma^{*-}}$ and $T_{MB\to\pi^+\Sigma^{*-}}$ are expanded
around the pole in the complex plane with the leading term of the Laurent series, which is written as 
\begin{eqnarray}
&&T_{MB^*\to\pi^+\Sigma^{*-}}=\frac{g_{\Lambda(1520)\to MB^*}g_{\Lambda(1520)\to\pi^+\Sigma^{*-}}}{\sqrt{s}-M_{\Lambda(1520)}+i\Gamma_{\Lambda(1520)}/2},\nonumber~~~~~\\
&&T_{MB\to\pi^+\Sigma^{*-}}=\frac{g_{\Lambda(1520)\to MB}g_{\Lambda(1520)\to\pi^+\Sigma^{*-}}}{\sqrt{s}-M_{\Lambda(1520)}+i\Gamma_{\Lambda(1520)}/2}.~~~~~\label{Eq:Tunit1520}
\end{eqnarray}
Here, $\Gamma_{\Lambda(1520)}$ is the total width of the $\Lambda(1520)$. Substituting Eq.~(\ref{Eq:Tunit1520}) into Eq.~(\ref{Eq:Amp1520}), the amplitude in Eq.~(\ref{Eq:Amp1520}) can be matched to the resonant process in Fig.~\ref{Fig:EffLab1520}, which is
\begin{eqnarray}
-it_{\Lambda(1520)}=\frac{g_{\Lambda(1520)\to\pi^+\Sigma^{*-}}g_{\Lambda(1520)\to\gamma\Lambda(\Sigma^0)} }{\sqrt{s}-M_{\Lambda(1520)}+i\Gamma_{\Lambda(1520)}/2}S^\dag\cdot\epsilon.~~~
\end{eqnarray}
This matching result gives the effective couplings $g_{\Lambda(1520)\to\gamma\Lambda}$ and $g_{\Lambda(1520)\to\gamma\Sigma^0}$ in terms of the loop functions $G_{MB^{(*)}}$ from Eq.~(\ref{Eq:Amp1520}) and the couplings $g_{\Lambda(1520)\to MB^{(*)}}$, namely,
\begin{eqnarray}
&&g_{\Lambda(1520)\to\gamma\Lambda(\Sigma^0)}=-e\frac{f_{\Delta\pi N}^*}{m_\pi}\sum_{MB^*}\xi_{MB^*}^{(s)}\nonumber\\
&&\qquad\qquad\qquad\qquad\times\left(G_{MB^*}^{(3b)}+G_{MB^*}^{(3c)}\right)g_{\Lambda(1520)\to MB^*}\nonumber\\
&&\qquad\qquad\qquad\qquad+\frac{e}{F_\phi}\sum_{MB}\xi_{MB}^{(d)}\nonumber\\
&&\qquad\qquad\qquad\qquad\times\left(G_{MB}^{(3d)}+G_{MB}^{(3e)}\right)g_{\Lambda(1520)\to MB},~~~~~~~
\end{eqnarray}
where the couplings $g_{\Lambda(1520)\to MB^*}$ and $g_{\Lambda(1520)\to MB}$ of the dynamically generated $\Lambda(1520)$ to the $S$-wave and $D$-wave coupled channels are collected in Table~\ref{Tab:Coup1520}. Note that these couplings are for isospin channels, and they should be multiplied by the corresponding Clebsch–Gordan coefficients for particle channels, as provided in Eqs.~(\ref{Eq:IsoCCMBs}) and (\ref{Eq:IsoCCMB}). Using the effective couplings $g_{\Lambda(1520)\to\gamma\Lambda(\Sigma^0)}$, one finds the partial decay width for the radiative decays $\Lambda(1520)\to\gamma\Lambda(\Sigma^0)$ as follows:
\begin{eqnarray}
\Gamma=\frac{|\vec{k}|}{3\pi}\left|g_{\Lambda(1520)\to\gamma\Lambda(\Sigma^0)}\right|^2\frac{m_{\Lambda(\Sigma^0)}}{M_{\Lambda(1520)}},\label{Eq:Gamma1520}
\end{eqnarray}
where $|\vec{k}|$ is the center-of-mass 3-momentum
of the photon in the $\Lambda(1520)$ rest frame. In this work, we fix the center-of-mass energy of the $\Lambda(1520)$ at $1520~{\rm MeV}$, which is
the real part of the $\Lambda(1520)$ pole position in the meson-baryon scattering amplitude. It should be emphasized that all loop functions are evaluated  on the physical sheet, i.e., at $\sqrt{s}\to \sqrt{s}+i\epsilon$. For the masses of octet baryons and pseudoscalar mesons, we take the PDG values~\cite{ParticleDataGroup:2024cfk}.

\begin{table}[htbp]
\caption{Couplings of the dynamically generated $\Lambda(1520)$ to the coupled channels in $S$ and $D$ waves~\cite{Roca:2006sz}.\label{Tab:Coup1520}}
    \centering
    \begin{tabular}{ccc}
    \hline
      \hline
       Channels  &  $g_{\Lambda(1520)\to MB^*}$ & $g_{\Lambda(1520)\to MB}$\\
       \hline
       $\pi\Sigma^*$ & $0.91$ & $-$\\
       
       $K\Xi^*$ & $-0.29$ & $-$\\
       
       $\bar{K}N$ & $-$ & $-0.54$\\
       
       $\pi\Sigma$ & $-$ & $-0.45$\\
       \hline
       \hline
    \end{tabular}
\end{table}
\begin{table*}[htbp]
\centering
\caption{\label{Tab:Result1520}Decomposition of each meson-baryon channel and diagram contributions to the effective couplings $g_{\Lambda(1520)\to\gamma\Lambda}$ and $g_{\Lambda(1520)\to\gamma\Sigma^0}$. The superscripts $(2b)$, $(2c)$, $(2d)$, and $(2e)$ denote the contributions from the diagrams $(b)$, $(c)$, $(d)
$, and $(e)$ in Fig.~\ref{Fig:Lab1520}, respectively. These effective couplings are given in units of $10^{-3}$.}
\scalebox{1}{
    \begin{tabular}{ccccccccccc}
      \hline
      \hline
      \multirow{2}{0.5cm}  &  & \multicolumn{4}{c}{$\Lambda(1520)\to\gamma\Lambda$} &  & \multicolumn{4}{c}{$\Lambda(1520)\to\gamma\Sigma^0$}\\
      \cline{3-6}\cline{8-11}
      &  & ~$g_{\Lambda(1520)\to\gamma\Lambda}^{(2b)}$~ & ~$g_{\Lambda(1520)\to\gamma\Lambda}^{(2b)+(2c)}$~ & ~$g_{\Lambda(1520)\to\gamma\Lambda}^{(2d)}$~ & ~$g_{\Lambda(1520)\to\gamma\Lambda}^{(2d)+(2e)}$~ & & ~$g_{\Lambda(1520)\to\gamma\Sigma^0}^{(2b)}$~ & ~$g_{\Lambda(1520)\to\gamma\Sigma^0}^{(2b)+(2c)}$~ & ~$g_{\Lambda(1520)\to\gamma\Sigma^0}^{(2d)}$~ & ~$g_{\Lambda(1520)\to\gamma\Sigma^0}^{(2d)+(2e)}$~\\
      \hline
      \multicolumn{2}{c}{$\pi^+\Sigma^{*-}$} & $-37.1$ & $-32.5$ & $-$ & $-$ &  & $11.3$ & $9.8$ & $-$ & $-$\\
      
      \multicolumn{2}{c}{$\pi^-\Sigma^{*+}$} & $37.1$ & $32.5$ & $-$ & $-$ &  & $11.3$ & $9.8$ & $-$ & $-$\\
      
      \multicolumn{2}{c}{$K^+\Xi^{*-}$} & $-2.3$ & $-1.7$ & $-$ & $-$ &  & $1.1$ & $0.9$ & $-$ & $-$\\

      \multicolumn{2}{c}{$K^-p$} & $-$ & $-$ & $-9.7-2.1i$ & $-5.9-1.3i$ &  & $-$ & $-$ & $2.8+0.6i$ & $1.7+0.4i$\\

      \multicolumn{2}{c}{$\pi^+\Sigma^-$} & $-$ & $-$ & $11.7+5.1i$ & $9.7+4.6i$ &  & $-$ & $-$ & $12.0+5.0i$ & $9.9+4.3i$\\

      \multicolumn{2}{c}{$\pi^-\Sigma^+$} & $-$ & $-$ & $-11.7-5.1i$ & $-9.7-4.6i$ &  & $-$ & $-$ & $12.0+5.0i$ & $9.9+4.3i$\\
      
      \hline
      \hline
    \end{tabular}}
\end{table*}

From Eq.~(\ref{Eq:Gamma1520}), the radiative decay widths and their ratio for the $\Lambda(1520)\to\gamma\Lambda$ and $\Lambda(1520)\to\gamma\Sigma^0$ are calculated to be
\begin{eqnarray} &&\Gamma(\Lambda(1520)\to\gamma\Lambda)=1.7{\rm ~keV},\nonumber\\
&&\Gamma(\Lambda(1520)\to\gamma\Sigma^0)=44.6{\rm ~keV},\nonumber\\
&&{\cal R}_{\Lambda(1520)}=\frac{\Gamma(\Lambda(1520)\to\gamma\Lambda)}{\Gamma(\Lambda(1520)\to\gamma\Sigma^0)}=0.04.\label{Eq:Pre1520}
\end{eqnarray}
While our prediction for the $\Gamma(\Lambda(1520)\to\gamma\Sigma^0)$ is in excellent agreement with the latest BESIII data~\cite{l6g2-2wg6}, that for the $\Gamma(\Lambda(1520)\to\gamma\Lambda)$ is significantly smaller than the CLAS measurement~\cite{CLAS:2005bgo}. Such a puzzling physical phenomenon should be checked by future experimental and theoretical studies.

In the previous work~\cite{Doring:2006ub}, the authors have studied the radiative decays of the $\Lambda(1520)$ using the chiral unitary approach. Unlike our findings, both of their predicted results, $\Gamma(\Lambda(1520)\to\gamma\Lambda)=2.5\sim4$~KeV and $\Gamma(\Lambda(1520)\to\gamma\Sigma^0)=71$~KeV, are inconsistent with the CLAS and the latest BESIII data~\cite{CLAS:2005bgo,l6g2-2wg6}. Two factors account for these differences. One is the numerical treatment of the $D$-wave contributions: we take $F_\phi=1.17F_\pi$, whereas Ref.~\cite{Doring:2006ub} uses $F_\phi=F_\pi$. The other is the inclusion of diagrams with a photon coupling to an intermediate baryon in our work, which are absent in Ref.~\cite{Doring:2006ub}. These differences explain why the predictions in Ref.~\cite{Doring:2006ub} are systematically larger than ours. To quantify individual effects of the two factors on the decay widths of the $\Lambda(1520)\to\gamma\Lambda(\Sigma^0)$, we consider two scenarios: Adopting $F_\phi=1.17F_\pi$ alone gives the $\Gamma(\Lambda(1520)\to\gamma\Lambda)_{\rm I}=4.0$~KeV and $\Gamma(\Lambda(1520)\to\gamma\Sigma^0)_{\rm I}=64.6$~KeV, whereas including only diagrams with a photon coupling to an intermediate baryon yields the $\Gamma(\Lambda(1520)\to\gamma\Lambda)_{\rm II}=2.1$~KeV and $\Gamma(\Lambda(1520)\to\gamma\Sigma^0)_{\rm II}=53.1$~KeV. It is found that the latter is already sufficient to explain the latest BESIII data of the $\Lambda(1520)\to\gamma\Sigma^0$~\cite{l6g2-2wg6}, and the addition of $F_\phi=1.17F_\pi$ further shifts the prediction toward the central value $46.8(66)(93){\rm ~keV}$ of the BESIII measurement.

In addition, one notes that our calculated ${\cal R}_{\Lambda(1520)}$ in Eq.~(\ref{Eq:Pre1520}) is less than 1, similar to the behavior predicted by flavor $SU(3)$ symmetry~\cite{Landsberg:1996gb,Mast:1968ltv}. Nevertheless, our value is much smaller than 1, while the $SU(3)$ symmetry predicts about $0.4$. To elucidate the underlying mechanism, we provide a detailed decomposition of the contributions to the effective coupling $g_{\Lambda(1520)\to\gamma\Lambda}$, itemized by each Feynman diagram and meson-baryon channel. These results are summarized in Table~\ref{Tab:Result1520}. As expected, the $S$-wave contributions to the radiative decays of the $\Lambda(1520)$ are all real, because the mass of this resonance lies below the meson-baryon threshold. For the $\Lambda(1520)\to\gamma\Lambda$ decay, Table~\ref{Tab:Result1520} shows that while the individual contribution from $S$-wave $\pi^+\Sigma^{*-}$ or $\pi^+\Sigma^{*-}$ is the largest, their sum vanishes exactly as a consequence of isospin symmetry. The same destructive interference also holds for the $\pi^+\Sigma^-$ and $\pi^-\Sigma^+$ channels in $D$ wave. These features collectively render the decay width of the $\Lambda(1520)\to\gamma\Lambda$ small. We also find that the $D$-wave contribution dominates the $\Lambda(1520)\to\gamma\Lambda$ decay, since the contribution from the $K^-p$ channel is significantly larger than that from the $K^+\Xi^{*-}$ channel. For the $\Lambda(1520)\to\gamma\Sigma^0$ decay, however, the contributions arising from each meson-baryon channel in $S$ and $D$ waves exhibit constructive interference with each other. This naturally gives rise to a larger decay width than that of the $\Lambda(1520)\to\gamma\Lambda$. Moreover, the $S$- and $D$-wave contributions are found to be roughly of the same order, with the $D$-wave contribution being marginally larger. As discussed above, the $D$-wave contributions are important and non-negligible in the study of the $\Lambda(1520)\to\gamma\Lambda(\Sigma^0)$. Additionally, an inspection of Table~\ref{Tab:Result1520} reveals that they noticeably affect the radiative decays of the $\Lambda(1520)$, despite the contributions from diagrams $(2c)$ and $(2e)$ being of relatively small magnitude. In particular, the consideration of diagrams $(2c)$ and $(2e)$ with a photon coupling to an intermediate baryon predicts smaller decay widths for the $\Lambda(1520)\to\gamma\Lambda(\Sigma^0)$, because these contributions are always opposite in sign to those of diagrams $(2b)$ and $(2d)$. This is one reason why our work can explain the latest BESIII data for the $\Lambda(1520)\to\gamma\Sigma^0$, whereas the previous U$\chi$PT work~\cite{Doring:2006ub} could not.

\begin{figure}[htbp]
  \centering
  \includegraphics[width=7.5cm]{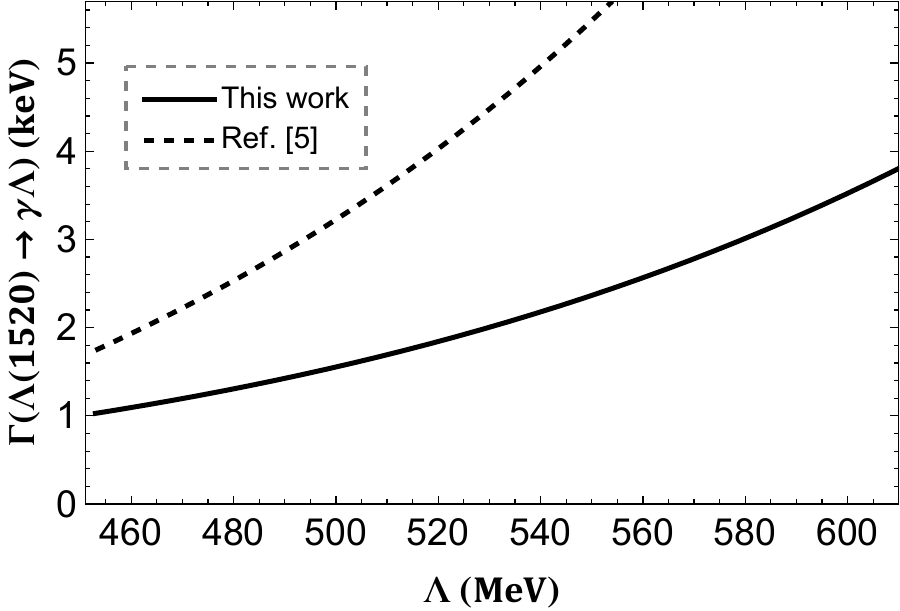}\\
  \caption{Decay width of the $\Lambda(1520)\to\gamma\Lambda$ as a function of the cutoff parameter $\Lambda$. The solid and
dashed lines represent the results of the present work and Ref.~\cite{Doring:2006ub}, respectively. }\label{Fig:CutoffLambda}
\end{figure}
\begin{figure}[htbp]
  \centering
  \includegraphics[width=7.5cm]{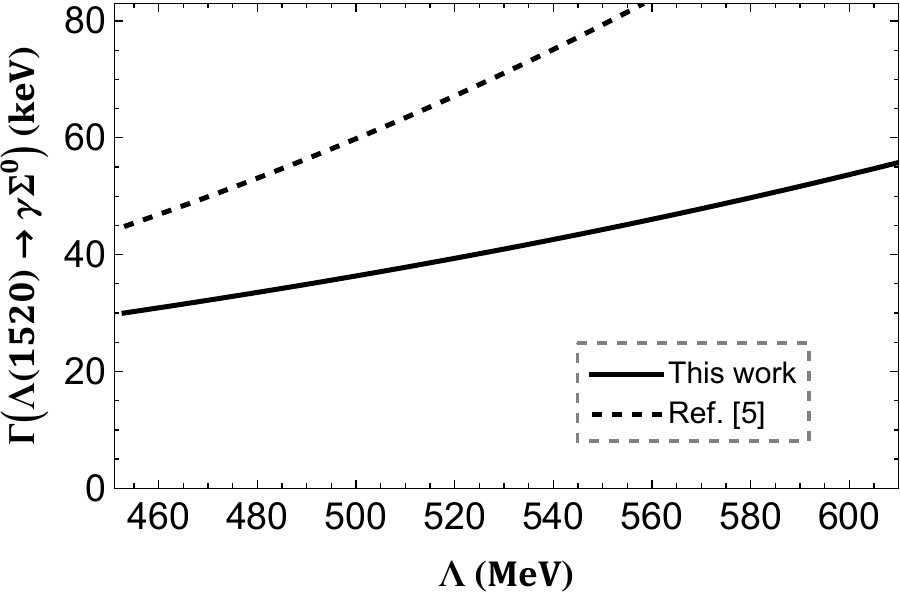}\\
  \caption{Same as Fig.~\ref{Fig:CutoffLambda} but for the $\Lambda(1520)\to\gamma\Sigma^0$.}\label{Fig:CutoffSigma0}
\end{figure}

Now, we discuss the cutoff dependence of the radiative decay widths of $\Lambda(1520)$. For this purpose, we have plotted Figs.~\ref{Fig:CutoffLambda} and \ref{Fig:CutoffSigma0} by using a common cutoff value for all $D$-wave meson-baryon channels. As shown in these two figures, in contrast to Ref.~\cite{Doring:2006ub}, our predicted decay widths exhibit substantially weaker cutoff dependence.  This is due to two reasons: (i) the $S$-wave contribution computed with the dimensional regularization scheme is cutoff-independent. (ii) for the $D$-wave, the inclusion of diagram $(2e)$ reduces the cutoff dependence, as its contribution has the opposite sign to that of diagram $(2d)$. Nevertheless, our predicted decay widths remain moderately sensitive to the cutoff parameter. This indicates that any substantial cutoff deviation from our adopted values, $\Lambda_{\bar{K}N}=507~{\rm MeV}$ and $\Lambda_{\pi\Sigma}=558~{\rm MeV}$, would inevitably shift the $\sqrt{s}$ value away from 1520 MeV. Consequently, the cutoff values need to be treated with due caution.

It should be recognized that the radiative decay amplitudes in this work are constructed using the
couplings $g_{\Lambda(1520)\to MB^{(*)}}$ of the $\Lambda(1520)$ to the coupled channels, extracted from the pole residues,
rather than the full coupled-channel amplitudes obtained from the Bethe-Salpeter equation. As a consequence, the radiative decay widths predicted in the present work inevitably suffer from uncertainties induced by the pole approximation. To examine whether these uncertainties could affect our predictions, we perform the following analysis. Since the radiative decay widths of $\Lambda(1520)$ are proportional to the squared modulus of the unitary matrix $T_{MB^{(*)}\to\pi\Sigma^*}$, their uncertainties are reflected by those of $|T_{MB^{(*)}\to\pi\Sigma^*}|^2$. Within the pole approximation, the values of $|T_{MB^{(*)}\to\pi\Sigma^*}|^2$ are computed from Eq.~(\ref{Eq:Tunit1520}) with $\sqrt{s}=1520~{\rm MeV}$. By a direct comparison between these values and those obtained from the full Bethe-Salpeter equation, we readily estimate the uncertainties for the $\Gamma(\Lambda(1520)\to\gamma\Lambda)$ and $\Gamma(\Lambda(1520)\to\gamma\Sigma^0)$ to be about 5\% and 7\%, respectively. These uncertainties are relatively small and therefore do not affect our previous conclusions.

\subsection{Discussion for the $\Lambda(1520)$ resonance component}

As discussed above, the purely dynamically generated meson–baryon picture within the chiral unitary approach reproduces the BESIII \(\Lambda(1520)\to\gamma\Sigma^0\) data well but substantially underestimates the CLAS \(\Lambda(1520)\to\gamma\Lambda\) measurement. This indicates that the \(\Lambda(1520)\) cannot be described purely as a meson-baryon molecular state and likely contains intrinsic three-quark component mixing. To deepen our understanding of the role of different components in the radiative decays of the $\Lambda(1520)$ resonance, we compare the results from the chiral unitary approach with those from quark models, where the $\Lambda(1520)$ is composed of three constituent quarks.

\begin{figure}[htbp]
  \centering
  \includegraphics[width=7cm]{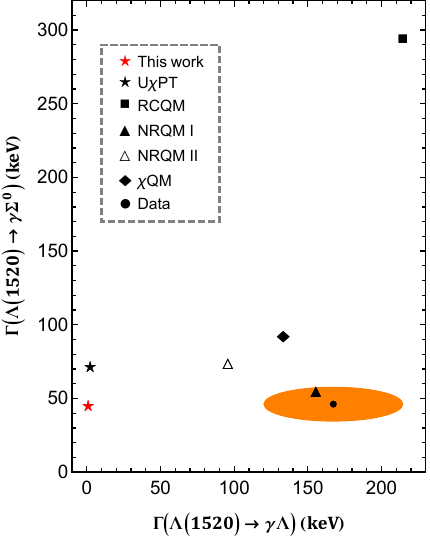}\\
  \caption{Decay widths of $\Lambda(1520)\to\gamma\Lambda$ and $\Lambda(1520)\to\gamma\Sigma^0$ obtained in various approaches. The black dots denote the results of the CLAS~\cite{CLAS:2005bgo} and BESIII~\cite{l6g2-2wg6} collaborations, and the orange contour corresponds to the $1\sigma$ confidence interval of the results. The red mark denotes our prediction. The others are taken from the U$\chi$PT of Ref.~\cite{Doring:2006ub}, the relativized constituent quark model~\cite{Warns:1990xi}~(RCQM), the non-relativistic quark models~(NRQM I~\cite{Kaxiras:1985zv} and NRQM II~\cite{Darewych:1983yw}), and the chiral quark model~\cite{Yu:2006sc}~($\chi$QM).}\label{Fig:Comparison}
\end{figure}

In Fig.~\ref{Fig:Comparison} we compare our predicted decay widths $\Gamma(\Lambda(1520)\to\gamma\Lambda)$ and $\Gamma(\Lambda(1520)\to\gamma\Sigma^0)$ with those obtained by other approaches, including the U$\chi$PT~\cite{Doring:2006ub}, the relativized constituent quark model~\cite{Warns:1990xi}~(RCQM), the non-relativistic quark models~(NRQM I~\cite{Kaxiras:1985zv} and NRQM II~\cite{Darewych:1983yw}), and the chiral quark model~\cite{Yu:2006sc}~($\chi$QM). We note that for the $\Lambda(1520)\to\gamma\Sigma^0$, all other quark model predictions fall outside the $1\sigma$ confidence interval of the latest BESIII data~\cite{l6g2-2wg6}, except for ours and that of NRQM I~\cite{Kaxiras:1985zv}. Nevertheless, our prediction $\Gamma(\Lambda(1520)\to\gamma\Sigma^0)=44.6{\rm ~keV}$ is significantly closer to the experimental central value $46.8~{\rm keV}$ than the result $55~{\rm~keV}$ from NRQM I. For the $\Lambda(1520)\to\gamma\Lambda$, the predictions from the chiral unitary approach are far from the CLAS measurement~\cite{CLAS:2005bgo}. By contrast, all quark model predictions are either consistent with the CLAS data or lie close to them. The above results warrant further verification through future experimental and theoretical studies.

To gain further insight into the internal structure of the $\Lambda(1520)$, we can estimate its composite nature by using the Weinberg compositeness~\cite{Aceti:2012dd,Aceti:2014wka,Liu:2026esv},
\begin{eqnarray}
\sum_i X_i=1-Z,\label{Eq:composite1}
\end{eqnarray}
with
\begin{eqnarray}
X_i=-{\rm Re}\left[g_i^2\left[\frac{dG_{ii}^{\rm II}(s)}{d\sqrt{s}}\right]_{\sqrt{s}=\sqrt{s_0}}\right],\label{Eq:composite2}
\end{eqnarray}
where $X_i$ represents the weight of each
meson-baryon channel. $Z$ is a measure of the presence of genuine $qqq$ components. The coupling strength $g_i$ of the dynamically generated $\Lambda(1520)$ to each meson-baryon channel in both the $S$- and $D$-waves is given in Table~\ref{Tab:Coup1520}. $\sqrt{s_0}$ denotes the complex pole in the second Riemann sheet, with its value fixed at $\sqrt{s_0}=(1520-i\Gamma_{\Lambda(1520)}/2)~{\rm MeV}$. $G_{ii}^{\rm II}(s)$ is the
analytic continuation of the loop function, which reads
\begin{eqnarray}
G_{ii}^{\rm II}(s) & = & 
\begin{cases} 
G_{ii}(s), & \text{Re}[\sqrt{s}] < (m_i + M_i), \\
G_{ii}^{\rm II}(s), & \text{Re}[\sqrt{s}] \geq (m_i + M_i),
\end{cases}
\end{eqnarray}
where $G_{ii}(s)$ is given in Eq.~(\ref{Eq:Gl}) and 
\begin{eqnarray}
G_{ii}^{\rm II}(s)=G_{ii}(s)+i\frac{2M_i}{4\pi\sqrt{s}}\frac{|\vec{q}|^{2l_{ii}+1}}{|\vec{q}|_{\rm on}^{2l_{ii}}}.
\end{eqnarray}
The weights corresponding to channels $\pi\Sigma^*$, $K\Xi^*$, $\bar{K}N$ and $\pi\Sigma$ are obtained from Eq.~(29) as $X_{\pi\Sigma^*}=0.13$, $X_{K\Xi^*}=0.00$, $X_{\bar{K}N}=0.48$ and $X_{\pi\Sigma}=0.24$, respectively. We find that the $\bar{K}N$ component plays a dominant role among all meson-baryon channels. Summing over all $X_i$ yields $1-Z=0.85$, which points to a pronounced meson-baryon component in the $\Lambda(1520)$ resonance, with the admixture of other genuine states constrained to less than 15\%.

\section{Conclusions}\label{SecIV}

In this work, we investigated the radiative decays $\Lambda(1520)\to\gamma\Lambda(\Sigma^0)$ using the chiral
unitary coupled-channel approach.  We have considered the combination of $S$- and $D$-wave coupling channels. The $S$- and $D$-wave Feynman diagrams are evaluated using the dimensional regularization and cut-off schemes, respectively. Furthermore, our results indicate a significant $D$-wave contribution, which dominates over the $S$-wave in the decay $\Gamma(\Lambda(1520)\to\gamma\Lambda)$, while in the decay $\Gamma(\Lambda(1520)\to\gamma\Sigma^0)$, its contribution is comparable to that of the $S$-wave. Interestingly, the chiral unitary approach can only explain the  BESIII data of the $\Gamma(\Lambda(1520)\to\gamma\Sigma^0)$, but not the $\Gamma(\Lambda(1520)\to\gamma\Lambda)$, which is much smaller than the CLAS experimental value~\cite{CLAS:2005bgo}. On the other hand, we found that the data of the $\Gamma(\Lambda(1520)\to\gamma\Lambda)$ can be well described by the nonrelativistic quark model~\cite{Kaxiras:1985zv} and the chiral quark model~\cite {Yu:2006sc}.

Further studies are required to clarify the internal structure of $\Lambda(1520)$. Future directions include: (i) Constructing mixed wave functions incorporating both three-quark and meson-baryon components; (ii) Studying strong and semileptonic decays of \(\Lambda(1520)\) to constrain its component composition; (iii) Awaiting more precise experimental measurements of \(\Gamma(\Lambda(1520)\to\gamma\Lambda)\) from BESIII and JLab to resolve the current theory-experiment tension.

\section{Acknowledgments}
This work is partly supported by the National Natural Science Foundation of China under Grant No. 12405091, the Natural Science Foundation of Guangxi province under Grant No. 2025GXNSFBA069314, and the Starting Research Fund from the Guangxi Normal University under Grant No.DC2300003299. L.S.G acknowledges support from the National Key R\&D Program of China under Grant No. 2023YFA1606703 and the National Natural Science Foundation of China under Grants No. W2543006 and  No. 12435007. J. X. L. acknowledges support from the National Natural Science Foundation of China under Grant No. 12522505. 

\bibliography{bib.bib}

\end{document}